\documentclass[onecolumn,tightenlines,nofootinbib,preprintnumbers,
amsmath,amssymb]{revtex4}

\usepackage[usenames,dvipsnames]{color}
\usepackage{caption2}
\usepackage{dcolumn}
\usepackage{graphicx}
\usepackage{subfigure}
\usepackage{epstopdf}
\usepackage{bm}
\usepackage{lscape}
\renewcommand{\arraystretch}{0.9}
\usepackage{setspace}

\begin{document}
\begin{spacing}{1.5}

\title{Direct CP violation in $\bar B_{s}  \rightarrow K^{+}K^{-} K^{+}K^{-}$ decay process induced by interferences of the intermediate vector particles}

\author{ Chang-Chang Zhang $^{1}$\footnote{Email: 1219765284@qq.com}, Gang L\"{u}$^{1}$\footnote{Corresponding author Email: ganglv66@sina.com}}

\affiliation{\small $^{1}$College of Physics, Henan University of Technology, Zhengzhou 450001, China\\
}

\begin{abstract}

We investigate CP violation in the decay process $\bar B_{s} \rightarrow V V \rightarrow K^{+}K^{-}K^{+}K^{-}$ within the framework of perturbative QCD, where V represents vector mesons $\phi$, $\rho$, and $\omega$. We analyze the mixing mechanism among $\phi-\rho^{0}-\omega$ and provide amplitudes for these decay processes. Moreover, we explore CP violation in the four-body decay process of $\bar{B}_{s}^{0} \rightarrow K^{+} K^{-} K^{+} K^{-}$ involving intermediate vector mesons and their mixing. Notably, significant CP violation is observed for specific two-vector meson intermediate states. Additionally, a substantial amount of CP violation arises from vector mixing when the invariant mass of $K^{+}K^{-}$ falls within a certain range. This study holds potential implications for future detection by the LHC experiment.

 \end{abstract}

\maketitle

\section{Introduction}
\label{sec:sample1}
The study of CP violation holds significant importance in the field of particle physics, and the non-zero weak phase within the Standard Model (SM) is responsible for CP violation \cite{Cabibbo:1963yz}.
However, theoretical calculations based on the weak phase of the SM yield results that differ significantly from experimental observations in certain decay processes, emphasizing the significance of CP violation as a means to explore new physics and mechanisms for CP violation.
In recent years, there has been a predominant focus on CP violation in two-body or three-body decay processes of B mesons, both theoretically and experimentally. With the increasing number of experimental measurements on four-body decay processes, the theoretical investigation into CP violation has garnered growing attention.

The experiments of CDF \cite{prl107-261802}
and LHCb \cite{plb713-369,prd90-052011,LHCb:2023exl} have published measurement results of CP violation for  $\bar B_{s} \rightarrow \phi \phi \rightarrow K^{+}K^{-}K^{+}K^{-}$,
which exhibit no discernible deviations from the predictions of the SM.
The ``real" triple product asymmetries (TPAs) from the four-body decay process of $\bar B_{s} \rightarrow (\pi \pi)(K \pi)$  is very small within the framework of perturbative QCD (PQCD), which is consistent with the predictions of the SM  \cite{Li:2021qiw}.
 Notably, significant CP violation is found in localized regions for both $B^{\pm} \rightarrow K^{\pm} \pi^{+} \pi^{-}$ and $B^{\pm} \rightarrow K^{\pm} K^{+} K^{-}$ decays by the LHCb experiment.
 The invariant mass spectra of $B^{\pm} \rightarrow K^{\pm} \pi^{+} \pi^{-}$ decays are studied within the region $0.08 \mathrm{GeV}^{2}/c^{4}<m_{\pi^{+}\pi^{-}}^{2}<0.66 \mathrm{GeV}^{2}/c^{4}$ and $m_{K^{\pm}\pi^{\mp}}^{2}<15 \mathrm{GeV}^{2}/c^{4}$, while the $B^{\pm} \rightarrow K^{\pm} K^{+} K^{-}$ decays are investigated in the range $1.2 \mathrm{GeV}^{2}/c^{4}<m_{K^+K^-low } ^{  2   }<  2.0\mathrm { GeV } ^ {  2   } / c ^ {    4   }$ and $m_{K^+K^-high } ^ {    2   }\textless15\mathrm { GeV } ^ {    2   }/c ^ {    4   }$
 \cite{LHCb:2012kja,LHCb:2012uja,Wang:2015ula,LHCb:2013ptu}.
 In recent years, an increasing number of analyses about precious measurements of the branching ratio and CP violation in the multiple body decay process have been carried out by BaBar \cite{BaBar:2014zli}, Belle II \cite{Bertacchi:2023jzv}, CLEO \cite{CLEO:2007vpk} and LHCb \cite{Aaijprl2013}, which provides a great platform to test the SM and search the new physical signals.
 Therefore, the investigation of CP violation in B meson four-body decay processes holds significant interest both theoretically and experimentally within specific localized regions.

We aim to utilize the PQCD method to calculate CP violation for the decay process of $\bar B_{s} \rightarrow K^{+}K^{-} K^{+}K^{-}$.
The Sudakov factor effectively suppress non-perturbative contributions  and absorb the non-perturbative part into universal hadronic wave functions in PQCD \cite{Ali:2007ff}.
PQCD method has been successfully applied to the non-leptonic two-body decay process of B meson  \cite{xiao2007xc}.
The corresponding two-body decay process of the B meson has been firmly established, followed by the development of PQCD method for various three-body and four-body decay processes, which can be regarded as quasi-two-body decay processes
\cite{Hua:2020usv, Zou:2020fax}.
Hence, we take the method of quasi-two-body decay process to calculate the CP violation of $\bar B_{s} \rightarrow K^{+}K^{-} K^{+}K^{-}$ process under the mixing mechanism of $\phi\rightarrow K^{+}K^{-}$, $\rho^{0}\rightarrow K^{+}K^{-}$ and $\omega\rightarrow K^{+}K^{-}$.
 By incorporating information on $K^{+}K^{-}$ production and taking into account the constraints imposed by isospin symmetry, quark model, and OZI rule, it becomes feasible to disentangle amplitudes with isospin $I=1$ and $ I=0$ components.
 The $\phi(1020)$ and $\omega$(782) match the isospin $I=0$ component. The $I=1$ component  derives from $\rho^{0}(770)$.

The three-particle mixing mechanism is based on the Vector Meson Dominance (VMD) model, where vector mesons are considered as propagators that interact with photons \cite{Nambu13,Kroll1967}.
The vector mesons of $\phi(1020)$ , $\omega$(782) and  $\rho^{0}(770)$
can be mixed as intermediate states.
The isospin field of intermediate states is transformed into physical field through the unitary matrix.
Hence, the physical amplitudes can be obtained to calculate
the CP violation.

We present our work in five distinct sections. In Section II, we perform a kinetic analysis of the four-body decay process. In Section III, we provide a comprehensive introduction to CP violation in the $\bar B_{s} \rightarrow K^{+}K^{-}K^{+}K^{-}$ decay process, including the mixing mechanism of $\phi$-$\omega$-$\rho^{0}$ vector particles and the contributions from resonance effects in $\bar B_{s} \rightarrow K^{+}K^{-}K^{+}K^{-}$ decay processes. In Section IV, we introduce the amplitude formalism within the framework of PQCD method, along with its fundamental functions and associated parameters. Furthermore, both the magnitude and integrated form of CP violation are evaluated. The data analysis results are also presented. The summary
and discussion can be found in Section V.

\section{ Analysis of the four-body decay kinematics}

The kinematics of the four-body decay of B-meson is much more complex than that of two-body decay. Both resonant and non-resonant contributions in the decay process can contribute to a large number of final state interactions \cite{MP,MT01,MT02,MT03,Grozin01,Grozin02}.
The decay process of $\bar B_{s} \rightarrow \phi\phi\rightarrow K^{+}K^{-}K^{+}K^{-}$ involves the sequential decay of a $\bar B_{s}$ meson into two $\phi$ mesons, followed by each individual $\phi$ meson decaying into a pair of $K^{+}K^{-}$ mesons.
The kinematics of four-body decay processes have been extensively examined by various researchers \cite{c1992,cly1993,g1990,a1965,g1979}.
\begin{figure}[h]
	\centering
	\includegraphics[height=6cm,width=12cm]{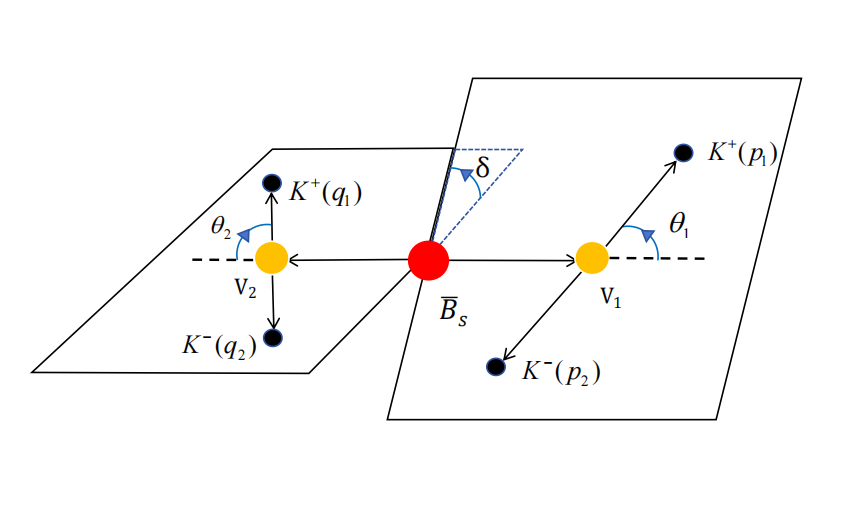}
	\caption{ The kinematic diagram of $\bar B_{s} \rightarrow V_{1}V_{2} \rightarrow K^{+}K^{-} K^{+}K^{-}$ process.}
	\label{fig1}
\end{figure}

The kinematic process of the four-body decay of $\bar{B}_s$ meson is illustrated in Fig. 1.
Five independent motion variables are present in the decay of the four-body $\bar B_{s}$ meson with initial spin 0.
The momenta of the initial $\bar B_{s}$ meson and the four final $K^{\pm}$ mesons are denoted as $P$, $p_{1}$, $p_{2}$, $q_{1}$ and $q_{2}$, respectively.
The above five variables are combined with Fig. 1 to yield the relations $s_1=(p_1+p_2)^2$, $s_2=(q_1+q_2)^2$.
Here, $\theta_1$ represents the angle between $p_{1}$ in the right-side $K^{+}K^{-}$ rest frame and the flight path of the right-side $K^{+}K^{-}$ system in the rest frame of $\bar B_{s}$ meson. Similarly, $\theta_2$ denotes the angle between $q_{1}$ in the left-side $K^{+}K^{-}$ rest frame and the flight path of the left-side $K^{+}K^{-}$ system in the rest frame of $\bar B_{s}$ meson. Additionally, $\delta$ signifies the inclination between the momentum-defining planes of both pairs of left and right side $K^{+} K^{-}$ mesons within $\bar B_{s}$ meson's rest system.
  The differential width of the four-body phase-space is defined as follows:
\begin{eqnarray}
d\Gamma=\frac{|{\cal M}|^2}{4(4\pi)^6 m_{\bar{B}_{s}}^{3}} X \beta_1 \beta_2 ds_1 ds_2 d\cos \theta_1 d\cos \theta_2 d\delta ~.
\end{eqnarray}
The expression $X=[(m_{\bar{B}_{s}}^2-s_1-s_2)^2/4-s_1s_2]^{1/2}$ is introduced, where $\beta_{\bf 1}=\lambda^{1/2}(s_1,m_{p_1}^2,m^{ 2}_{p_2})/s_1$ and $\beta_{\bf 2}=\lambda^{1/2}(s_2,m_{q_1}^2,m^{  2}_{q_2})/s_2$, with $\lambda(a,b,c)=a^2+b^2+c^2-2ab-2bc-2ca$. Here, $|{\cal M}|^2$ represents the squared amplitude obtained by summing over the spins \cite{Hsiao:2017nga}.
For the integration, the allowed ranges of $(s_1,s_2)$ and $(\theta_{\bf 1},\theta_{\bf 2},\delta)$
are given by
\begin{eqnarray}
	(m_{p_{1}}+m_{p_{2}})^2\leq &\,s_1\,&\leq (m_{\bar{B}_{s}}-\sqrt{s_2})^2\,,\nonumber\\
	(m_{q_1}+m_{q_2})^2\leq &\,s_2\,&\leq (m_{\bar{B}_s}-m_{p_1}-m_{p_2})^2\,,\nonumber\\
	0\leq \theta_{\bf 1,2}\leq \pi&\,,&0\leq \delta \leq 2\pi\,.
\end{eqnarray}



In three-body decays, the final state typically involves the decay of intermediate particles into two hadrons through strong interactions, resulting in the formation of a three-body final state.
This approach is also applied to four-body decays of $\bar{B}_s$ mesons, incorporating contributions from quasi-two-body decay processes.
We investigate the decay process $\bar B_{s} \rightarrow V_1$$V_2 \rightarrow K^{+}K^{-} K^{+}K^{-}$, where $V_1$ and $V_2$ represent two intermediate vector particles that further decay into two hadrons, namely $K^{+}K^{-}$.
 We employ the factorization relation, commonly referred to as the narrow width approximation (NWA), to decompose this process into a continuous two-body decay:
\begin{eqnarray}
	\Gamma\left(\bar B_{s} \rightarrow V_1V_2 \rightarrow K^{+}K^{-} K^{+}K^{-}\right)   =
	\Gamma\left(B \rightarrow V_1V_2\right) \mathcal{B}\left(V_1 \rightarrow  K^{+}K^{-}\right)\mathcal{B}\left(V_2 \rightarrow  K^{+}K^{-}\right),
\end{eqnarray}
where $\Gamma$ and B  represent the width and branching ratio of the decay process, respectively. Eq.(3) can be safely applied for quasi-two-body decay processes with small width.
The narrow-width approximation correction is required on this basis.

For the three-body decay process, R denotes an intermediate resonant state, while $P_{1}$, $P_{2}$, and $P_{3}$ represent pseudoscalar mesons.
 The modification factor $\eta_{R}$ for quasi-two-body decays, given by $\eta_{R} \Gamma\left( B \rightarrow RP_{3} \rightarrow P_{1}P_{2} P_{3}\right) =\Gamma\left(B \rightarrow RP_{3}\right) \mathcal{B}\left(R \rightarrow  P_{1}P_{2}\right)$, is determined to be less than $10\%$ when applying the QCD factorization method for correction calculations \cite{chenghaiyang2021prd,chenghaiyang2021plb}. Hence,
the parameter $\eta_{R}$ is introduced as a constant in  $\eta_{R} \Gamma\left(\bar B_{s} \rightarrow V_1V_2 \rightarrow K^{+}K^{-} K^{+}K^{-}\right)   =\Gamma\left(B \rightarrow V_1V_2\right) \mathcal{B}\left(V_1 \rightarrow  K^{+}K^{-}\right)\mathcal{B}\left(V_2 \rightarrow  K^{+}K^{-}\right)$. When calculating CP violation, this constant can be eliminated without affecting our final results.

\section{CP violation in $\bar B_{s} \rightarrow \phi$ ($\rho^{0}$, $\omega$) $ \phi$ ($\rho^{0}$, $\omega$) $  \rightarrow K^{+}K^{-} K^{+}K^{-}$ decay process }
\subsection{The mechanism of mixing three vector mesons}

The positrons and electrons  annihilate into photons and then they are polarized in a vacuum to form the mesons of $\phi (1020)$, $\rho^0(770)$ and $\omega(782)$, which can also decay into $K^{+} K^{-}$ pair in the VMD model \cite{PMplb1981,Achasov2016}.
 Since the intermediate state particle is an unphysical state, we need to convert it into a physical field from an isospin field through the matrix R \cite{Lu:2022rdi}. Then we can obtain the physical state of $\phi$, $\rho^{0}$ and $\omega$. What deserved to mentioned is that there is no $\phi- \rho^{0}- \omega$ mixing in the physical state \cite{Lv2023epj}.
 The physical states $\phi- \rho^{0}- \omega$
 can be expressed as linear combinations of the isospin states
 $\phi_{I}- \rho^{0}_{I}- \omega_{I}$.
 This relationship can be represented by the following matrix:

\begin{equation}
	\left (
	\begin{array}{lllll}
		\rho^0\\[0.5cm]
		\omega\\[0.5cm]
		\phi
	\end{array}
	\right )
	=
	R(s)
	\left (
	\begin{array}{lll}
		\rho^0_I\\[0.5cm]
		\omega_I\\[0.5cm]
		\phi_I
	\end{array}
	\right )
 =
	\left (
	\begin{array}{lll}
		<\rho_{I}|\rho> & \hspace{0.3cm} <\omega_{I}|\rho>  &\hspace{0.3cm}<\phi_{I}|\rho>\\[0.5cm]
		<\rho_{I}|\omega> &  \hspace{0.3cm}<\omega_{I}|\omega>&\hspace{0.3cm}<\phi_{I}|\omega>\\[0.5cm]
		<\rho_{I}|\phi>&\hspace{0.3cm} <\omega_{I}|\phi> & \hspace{0.3cm} <\phi_{I}|\phi>
	\end{array}
	\right )
\left (
	\begin{array}{lll}
		\rho^0_I\\[0.5cm]
		\omega_I\\[0.5cm]
		\phi_I
	\end{array}
	\right )
	\label{L1}.
\end{equation}
\noindent

The off-diagonal elements of R present the information of $\phi- \rho^{0}- \omega$ mixing.
Based on the isospin representation of $\phi_{I}$, $\rho_{I}$ and $\omega_{I}$, the isospin vector $|I,I_{3}>$ can be constructed,
where $I_3$ denotes the third component of isospin.
We use the notation $F_{V_iV_j}$ to denote the information mixing, where $V_i$ and $V_j$ represent one of the three vector particles. Then,
the transformation matrix R can then be converted as follows:
\begin{equation}
 R=\left(\begin{array}{ccc}
1 &  \hspace{0.3cm}-F_{\rho \omega}(s) &\hspace{0.3cm} -F_{\rho \phi}(s) \\
\\
F_{\rho \omega}(s) &\hspace{0.3cm} 1 &\hspace{0.3cm} -F_{\omega \phi}(s) \\
\\
F_{\rho \phi}(s) &\hspace{0.3cm} F_{\omega \phi}(s) &\hspace{0.3cm} 1
\end{array}\right).
\end{equation}

From the translation of the two representations, the physical states can be written as $\phi=F_{\rho\phi  }(s) \rho_{I}^{0}+F_{\omega \phi}(s) \omega_{I}+\phi_{I}$, $\omega=F_{ \rho\omega }(s) \rho_{I}^{0}+\omega_{I} -F_{\omega \phi}(s) \phi_{I}$ and $\rho^{0}=\rho_{I}^{0}-F_{\rho\omega }(s) \omega_{I}-F_{\rho\phi }(s) \phi_{I}$.
And the relationships between the physical states with the mixing parameters are $F_{\rho \omega}=\frac{\Pi_{\rho \omega}}{s_{\rho}-s_{\omega}}$,  $F_{\rho \phi}=\frac{\Pi_{\rho \phi}}{s_{\rho}-s_{\phi}}$ and $F_{\omega \phi}=\frac{\Pi_{\omega \phi}}{s_{\omega}-s_{\phi}}$, where $ F_{V_{i} V_{j}}$=$-F_{V_{j} V_{i}}$.
The inverse propagator of vector meson is defined as $s_{V}=s-m_{V}^{2}+\mathrm{i} m_{V} \Gamma_{V}$ ,
where $V = \phi, \rho$, or $\omega$.
$m_V$ and $\Gamma_{V}$ are the mass and decay rate of the vector mesons, respectively. $\sqrt{s}$ denotes the invariant mass of the $K^{+}K^{-}$ pair.

The momentum dependence of the mixing parameter $\Pi_{V_{i} V_{j}}$  is introduced, which is s dependent.
The precise determination of the mixing parameter $\Pi_{\rho \omega }=-4470 \pm 250 \pm 160-i(5800 \pm 2000 \pm 1100)  \mathrm{MeV}^{2}$ near the $\rho$ meson is accomplished. Similarly, the mixing parameter $\Pi_{\omega \phi}=19000+i(2500 \pm 300) \mathrm{MeV}^{2}$ near the $\phi$ meson is obtained \cite{CE2009,CE2011,Lu:2018fqe}. Additionally, Achasov et al. determine the mixing parameter $\Pi_{\phi\rho}=720 \pm 180 -i(870 \pm 320)\mathrm{MeV}^{2}$ in close proximity to the $\phi$ meson \cite{Achasov:1999wr}. Then we define
\begin{eqnarray}
\widetilde{\Pi}_{\rho\omega}=\frac{s_{\rho}\Pi_{\rho\omega}}{s_{\rho}-s_{\omega}},
~~\widetilde{\Pi}_{\rho\phi}=\frac{s_{\rho}\Pi_{\rho\phi}}{s_{\rho}-s_{\phi}},
~~\widetilde{\Pi}_{\phi\omega}=\frac{s_{\phi}\Pi_{\phi\omega}}{s_{\phi}-s_{\omega}}.
\end{eqnarray}

\subsection{The resonance effect }

\label{sec:spectra}
\begin{figure}[h]
	\centering
	\includegraphics[height=8cm,width=16cm]{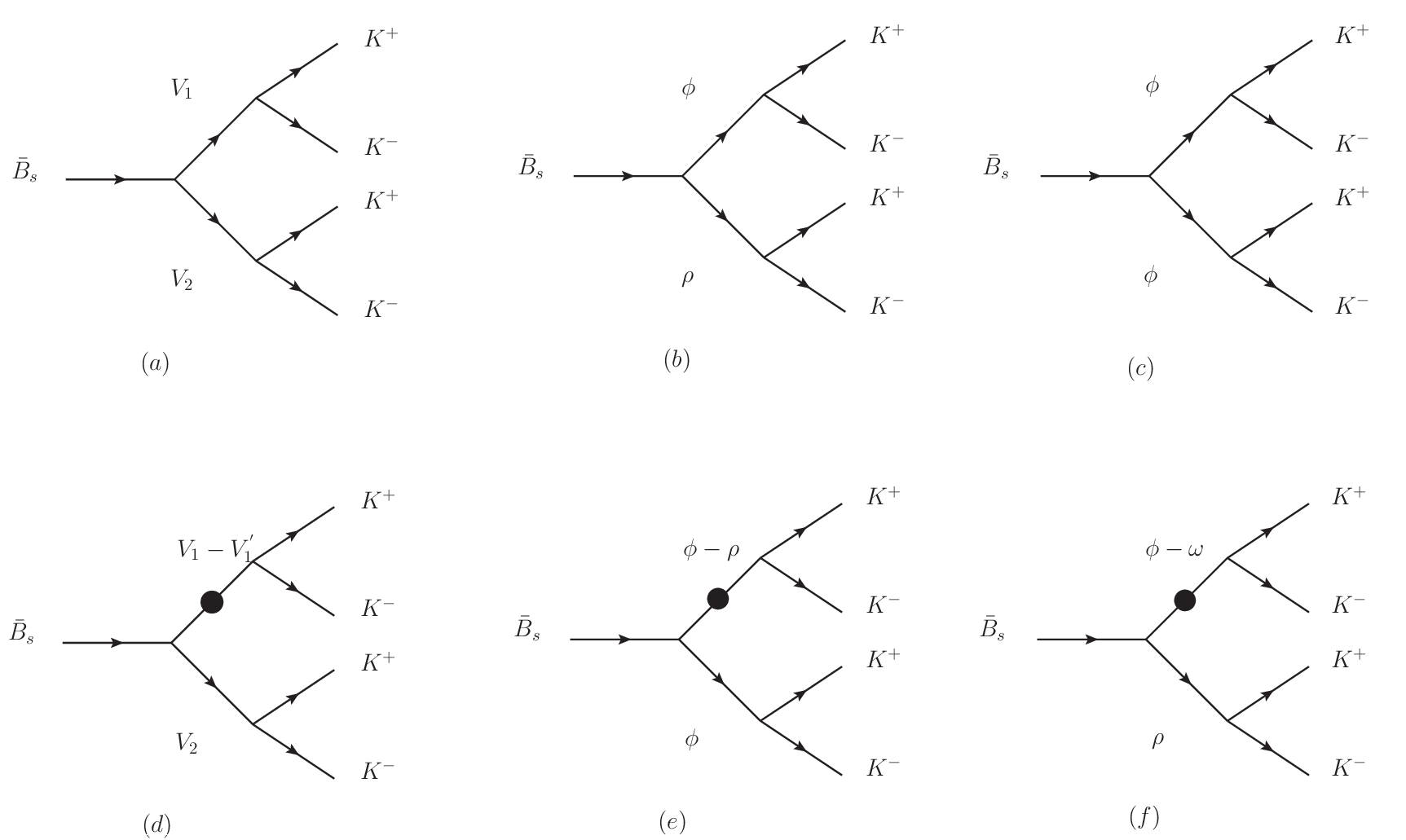}
	\caption{ The decay diagrams of $\bar B_{s} \rightarrow \phi $ $(\rho^{0}$, $\omega)$ $\phi $ $(\rho^{0}$, $\omega)$ $ \rightarrow K^{+}K^{-}K^{+}K^{-}$ process.}
	\label{fig2}
\end{figure}

We present the decay diagrams (a)-(f) of the $\bar B_{s} \rightarrow \phi$ $(\rho^{0}$, $\omega)$ $ \phi$ $(\rho^{0}$, $\omega)$ $ \rightarrow K^{+} K^{-} K^{+}K^{-}$ process in Fig. \ref{fig2} to provide a more comprehensive understanding of the mixing mechanism.
The decay process shown in (a) of Fig. 2 corresponds to direct decay modes, where the vector particles $V_1$ and $V_2$ represent one of the three possible vector mesons: $\phi$, $\rho^0$ and $\omega$, respectively, resulting in the production of $K^{+} K^{-}$.
Diagram (b) and diagram (c) serve as illustrative examples of diagram (a) in Fig. 2.
By pairwise pairing the three particles, diagram (a)  shows the distinct decay contributions of the six intermediate states leading to a final state of $K^{+}K^{-} K^{+}K^{-}$.
Compared to the direct decay processes depicted in diagrams (a), (b) and (c) of Fig.$~$\ref{fig2}, the $K^{+} K^{-}$ pair can also be produced through the mixing mechanism.
The black dots in the Fig. 2 represent the resonance effect between two vector mesons, described by the mixing parameter $\Pi_{V_{i} V_{j}}$.
 Although the contribution from this mixing mechanism is relatively small compared to other diagrams in Fig.$~$\ref{fig2} , it must still be taken into consideration.
Any vector particle may potentially undergo mixing with two other vector particles that are different from itself, as shown in  (d) of Fig.$~$\ref{fig2}.
Consider the example illustrated in (e) and (f) of Fig.$~$\ref{fig2}. If the intermediate state of the decay process comprises two identical particles, each of these particles can undergo mixing with another particle, thereby contributing to the decay.
Considering the factors associated with indistinguishable particles, this contribution is doubled.
The simultaneous mixing of two mesons in an intermediate state involves two mixing parameters, which are higher-order terms that have been overlooked.
Combining the contributions from the decay process shown in Fig. 2, we can obtain the form of the total amplitude as follows:
\begin{eqnarray}
\begin{split}
\left\langle K^{+} K^{-} K^{+} K^{-}\left|A\right| \bar B_{s}\right\rangle=
&
\frac{g_{\phi}^{2}}{s_{\phi}^{2}}A_{\phi\phi}
+\frac{2g_{\rho}g_{\phi}}{s_{\rho}s_{\phi}^{2}}\widetilde{\Pi}_{\rho\phi}A_{\phi\phi}
+\frac{2 g_{\omega}g_{\phi}}{s_{\omega}s_{\phi}^{2}}\widetilde{\Pi}_{\omega\phi}A_{\phi\phi}
+\frac{g_{\rho}^{2}}{s_{\rho}^{2}}A_{\rho\rho}
+\frac{2g_{\phi}g_{\rho}}{s_{\phi}s_{\rho}^{2}}\widetilde{\Pi}_{\phi\rho}A_{\rho\rho}
\\
&
+\frac{2g_{\omega}g_{\rho}}{s_{\omega}s_{\rho}^{2}}\widetilde{\Pi}_{\omega\rho}A_{\rho\rho}
+\frac{g_{\omega}^{2}}{s_{\omega}^{2}}A_{\omega\omega}
+\frac{2g_{\phi}g_{\omega}}{s_{\phi}s_{\omega}^{2}}\widetilde{\Pi}_{\phi\omega}A_{\omega\omega}
+\frac{2g_{\rho}g_{\omega}}{s_{\rho}s_{\omega}^{2}}\widetilde{\Pi}_{\rho\omega}A_{\omega\omega}
\\
&
+\frac{g_{\phi}g_{\rho}}{s_{\phi}s_{\rho}}A_{\phi\rho}
+\frac{g_{\phi}^{2}}{s_{\phi}^{2}s_{\rho}}\widetilde{\Pi}_{\phi\rho}A_{\phi\rho}
+\frac{g_{\phi}g_{\omega}}{s_{\phi}s_{\omega}s_{\rho}}\widetilde{\Pi}_{\omega\rho}A_{\phi\rho}
+\frac{g_{\rho}^{2}}{s_{\phi}s_{\rho}^{2}}\widetilde{\Pi}_{\rho\phi}A_{\phi\rho}
+\frac{g_{\rho}g_{\omega}}{s_{\phi}s_{\rho}s_{\omega}}\widetilde{\Pi}_{\omega\phi}A_{\phi\rho}
\\
&
+\frac{g_{\phi}g_{\omega}}{s_{\phi}s_{\omega}}A_{\phi\omega}
+\frac{g_{\phi}^{2}}{s_{\phi}^{2}s_{\omega}}\widetilde{\Pi}_{\phi\omega}A_{\phi\omega}
+\frac{g_{\phi}g_{\rho}}{s_{\phi}s_{\rho}s_{\omega}}\widetilde{\Pi}_{\rho\omega}A_{\phi\omega}
+\frac{g_{\omega}^{2}}{s_{\phi}s_{\omega}^{2}}\widetilde{\Pi}_{\omega\phi}A_{\phi\omega}
+\frac{g_{\omega}g_{\rho}}{s_{\phi}s_{\omega}s_{\rho}}\widetilde{\Pi}_{\rho\phi}A_{\phi\omega}
\\
&
+\frac{g_{\rho}g_{\omega}}{s_{\rho}s_{\omega}}A_{\rho\omega}
+\frac{g_{\rho}^{2}}{s_{\rho}^{2}s_{\omega}}\widetilde{\Pi}_{\rho\omega}A_{\rho\omega}
+\frac{g_{\rho}g_{\phi}}{s_{\rho}s_{\omega}s_{\phi}}\widetilde{\Pi}_{\phi\omega}A_{\rho\omega}
+\frac{g_{\omega}^{2}}{s_{\rho}s_{\omega}^{2}}\widetilde{\Pi}_{\omega\rho}A_{\rho\omega}
+\frac{g_{\omega}g_{\phi}}{s_{\rho}s_{\omega}s_{\phi}}\widetilde{\Pi}_{\phi\rho}A_{\rho\omega}.
\label{Htr}
\end{split}
\end{eqnarray}

The  amplitudes of $A_{\rho \rho}$, $A_{\omega \omega}$,  $A_{\phi \phi}$, $A_{\rho \omega}$, $A_{\omega \phi}$ and $A_{\phi \rho}$ correspond to the decay processes of $\bar B_s \rightarrow \rho^0 \rho^0$, $\bar B_s \rightarrow \omega \omega$,  $\bar B_s \rightarrow \phi \phi$,
$\bar B_s \rightarrow \rho^0 \omega$, $\bar B_s \rightarrow \omega \phi$ and $\bar B_s \rightarrow \phi \rho^0$, respectively. Here, $s_V$ represents the inverse propagator of the vector meson V  \cite{ Chen:1999nxa, Wolfe:2009ts, Wolfe:2010gf}.
Moreover, $g_{V}$ represents the coupling constant derived from the decay process of $ V \rightarrow K^{+} K^{-}$ and can be expressed as $\sqrt{2}g_{{\rho}k^{+} k^{-}}=\sqrt{2}g_{\omega k^{+} k^{-}}=-g_{\phi k^{+} k^{-}}=4.54$ \cite{Bruch:2004py}.

The amplitude of the $\bar{B}_{s} \rightarrow \phi $ ($\rho^{0}, \omega$) $ \phi $ ($\rho^{0}, \omega$) $  \rightarrow K^{+}K^{-}  K^{+}K^{-} $ decay channel can be characterized  as follows:
\begin{equation}
	A=\left \langle K^{+}K^{-} K^{+}K^{-}\left | H^{T} \right | \bar{B}_{s} \right \rangle +\left \langle K^{+}K^{-} K^{+}K^{-}\left | H^{P} \right | \bar{B}_{s} \right \rangle,
\end{equation}
where $\left \langle K^{+}K^{-} K^{+}K^{-}\left | H^{T} \right | \bar{B}_{s} \right \rangle $
  and $\left \langle K^{+}K^{-} K^{+}K^{-}\left | H^{P} \right | \bar{B}_{s} \right \rangle $
  represent  tree-level contributions and
  the amplitudes associated with penguin-level, respectively.

 The differential parameter for CP violation can be expressed as follows:
\begin{equation}
	\label{cp31}
	A_{CP}=\frac{\left| A \right|^2-\left| \overline{A} \right|^2}{\left| A \right|^2+\left| \overline{A} \right|^2}.
\end{equation}

In this work, we will consider the resonant and non-resonant contributions from a specific region $\Omega$. By integrating the numerator and denominator of $A_{CP}$ over this region, we obtain the local integrated CP violation, which can be measured experimentally. It takes the following form:
\begin{equation}
	A_{C P}^{\Omega}=\frac{\int \mathrm{d} \Omega\left(|A|^{2}-|\overline{A}|^{2}\right)}{\int  \mathrm{d} \Omega\left(|A|^{2}+|\overline{A}|^{2}\right)}.
\end{equation}

\section{\label{sec:cpv1}The amplitudes of quasi-two-body decay process within the framework of perturbative QCD }

\subsection{Formulation of calculations}

The PQCD method is a primary approach for investigating the non-leptonic decay of B meson. In this method, it is postulated that the primary process in B meson decay involves hard gluon exchange. The hard component is isolated and analyzed using perturbation theory, while the non-perturbative part is incorporated into the universal hadron wave function. Similar to the analysis of three-body decay process, the calculation of quasi-two-body decay process  involves defining an intermediate decay state. In line with the ``color transparency mechanism" within the B-meson initial system, the decay of  B meson results in light quarks move at high velocities and spectator quarks at rest. During the decay process, a hard gluon is exchanged between the spectator quark and the weakly interacting tetraquark operator, causing the spectator quark to accelerate. One of the light quarks combines with a spectator quark possessing higher kinetic energy to generate a swiftly moving final meson. The two ``final state" vector mesons subsequently decay into  $K^{+}K^{-}$, ultimately yielding four final-state particles. By clarifying the decay concept, we present the decay amplitude specific to this process.

\begin{figure}[h]
	\centering
	\includegraphics[height=6cm,width=18cm]{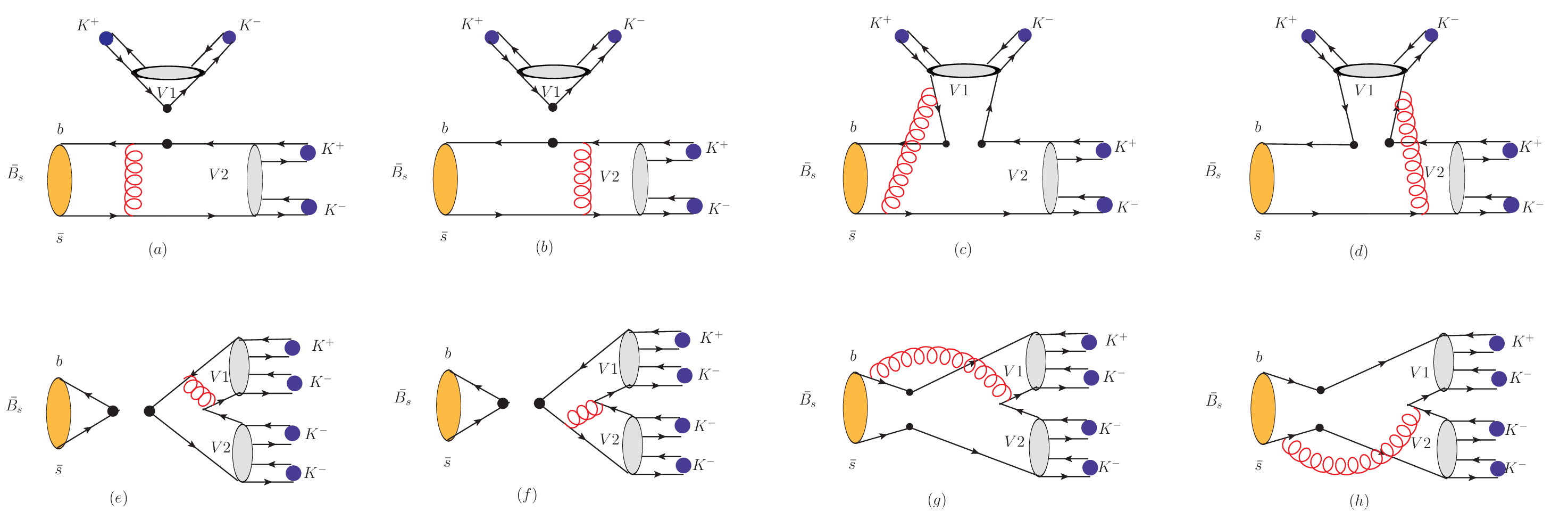}
	\caption{ The Feynman diagrams of $\bar B_{s} \rightarrow V_{1}V_{2} \rightarrow K^{+}K^{-} K^{+}K^{-}$ process.}
	\label{fig3}
\end{figure}

In Fig.$~$\ref{fig3}, (a)-(d) represent Feynman diagrams illustrating the emission contributions, including four possible diagrams with the insertion of four quark operators while (e)-(h) represent Feynman diagrams illustrating the annihilation
 contributions, including four possible diagrams with the insertion of four quark operators. By employing the quasi-two-body decay method, the total amplitude of  $\bar B_{s} \rightarrow V_1V_2$ $\rightarrow K^{+}K^{-} K^{+}K^{-} $ is composed of two components: $\bar B_{s} \rightarrow V_1V_2 $  and $V_1V_2$ $\rightarrow K^{+}K^{-} K^{+}K^{-} $.

When calculating the $\bar B_{s} \rightarrow V_1V_2 $ decay amplitude within the PQCD framework, it is essential to consider the angular distribution and utilize light-cone coordinates to define the four-momentum of the B meson and two vector mesons. By analyzing the Lorentz structure, the amplitude can be decomposed as follows:
\begin{equation}
A^{(\sigma)}_1=\epsilon^*_{1\mu}\epsilon^*_{2\nu}(a g^{\mu \nu}+\frac{b}{M_{1} M_{2}} P^{\mu} P^{\nu}+\frac{i c}{M_{1} M_{2}} \epsilon^{ \mu \nu \alpha \beta} P_{1 \alpha} P_{2 \beta}),
\end{equation}
where $\sigma$ represents the three polarizations of vector mesons, namely longitudinal (L), transverse (N), and perpendicular (T).
The amplitudes are characterized by the polarization states of these vector mesons when the $\bar B_{s}$ meson decays into two vector mesons.
Then we introduce another set of equivalent helicity amplitudes of
 $H_{0}=M^{2}_{\bar{B}_{s}} {\cal A}_L$, $H_{\pm}= M^{2}_{\bar{B}_{s}} {\cal A}_{N} \mp  M_{1} M_{2} \sqrt{r^{'2}-1}{\cal A}_{T}$,
where $r^{'2}$ is equal to $P_{1}\cdot P_{2}/(M_{1}M_{2})= M^{2}_{\bar{B}_{s}}/2(M_{1}M_{2})$.
 $P_{1}$, $P_{2}$, $M_{1}$ and $M_{2}$  represent the momentum and mass of two intermediate state vector particles, respectively \cite{Chen:2002pz}.

Next, we define the longitudinal  $H_{0}$ and  transverse  helicity amplitudes of $H_{\pm}$. Then we get
\begin{equation}
		\sum_{\sigma}{ A_1}^{(\sigma)\dagger }{ A_1^{(\sigma)}}=|H_{0}|^{2}+|H_{+}|^{2} + | H_{-}|^{2}.
\end{equation}

 Similarly, in the decay process of $V \rightarrow K^{+}K^{-}$, we can express $A_{V_1 \rightarrow K^{-} K^{+}}^{\lambda}=g_{V_1}\epsilon(\lambda)\left(p_1-p_2\right)$ and $A_{V_2 \rightarrow K^{-} K^{+}}^{\gamma}=g_{V_2}\epsilon(\gamma)\left(q_1-q_2\right)$,  where $\epsilon$ denotes the polarization vector of vector meson. $g_V$ represents an effective coupling constant for $V \rightarrow K^{+}K^{-}$.

Therefore  we employ the quasi-two-body decay method to calculate amplitude of four-body decay and then we obtain the complete  amplitude form for the decay process of  $\bar B_{s}^0\to V_{1} V_{2}  \rightarrow K^{+}K^{-}K^{+}K^{-}$:
\begin{eqnarray}
	\langle K^{+}K^{-} K^{+}K^{-}\left | H \right | \bar{B}_{s}  \rangle
	&=&   \langle K^{+}K^{-} \left | {\cal H}_{eff} \right | V_1
	 \rangle \langle K^{+}K^{-}  | {\cal H}_{eff}  | V_2  \rangle
	 \langle V_1V_2 \left | {\cal H}_{eff} \right | \bar{B}_{s} \rangle \nonumber\\
	&=&\frac{g_{V_1}\epsilon(\lambda)\left(p_1-p_2\right)
		g_{V_2}\epsilon(\gamma)\left(q_1-q_2\right) }{s_{V_1} s_{V_2}}\nonumber\\
		&&\epsilon^*_{1\mu}\epsilon^*_{2\nu}(a g^{\mu \nu}+\frac{b}{M_{1} M_{2}} P^{\mu} P^{\nu}+\frac{i c}{M_{1} M_{2}} \epsilon_{ \mu \nu \alpha \beta} P^{ \alpha}_1 P^{ \beta}_2).
\end{eqnarray}

Utilizing the principles of four-body kinematics, we provide the amplitude forms of the $\bar B_{s}^0\to V_{1} V_{2}  \rightarrow K^{+}K^{-}K^{+}K^{-}$ decay process without
the mixing of vector mesons as follows:
\begin{eqnarray}
	\sqrt 2 A^i(\bar B_{s}^0\to \phi(\phi  \rightarrow K^{+}K^{-})\phi(\phi  \rightarrow K^{+}K^{-})) &=& - \frac{2G_Fg_{\phi}\epsilon(\lambda)\left(p_1-p_2\right)
		g_{\phi}\epsilon(\gamma)\left(q_1-q_2\right)  }{\sqrt{2}s_{\phi} s_\phi}
	V_{tb}V_{ts}^{*}\nonumber
	\\
	&&\times  \Bigg\{ f_\phi F_{B_s\to
		\phi}^{LL,i}\left[a_3+a_{4}+a_5
	-\frac{1}{2}a_7 -\frac{1}{2}a_9-\frac{1}{2}a_{10}
	\right]\nonumber
	\\
	&&+ M_{B_s\to \phi}^{LL,i} \left[C_3+C_{4}-\frac{1}{2}C_9-\frac{1}{2}C_{10}\right]
	-M_{B_s\to \phi}^{LR,i} \left[C_5-\frac{1}{2}C_{7}\right]
	\nonumber
	\\
	&&	- M_{B_s\to \phi}^{SP,i}
	\left[C_6-\frac{1}{2}C_{8}\right]
	+ f_{B_s} F_{ann}^{LL,i}\left[a_3+a_{4}-\frac{1}{2}a_9-\frac{1}{2}a_{10}\right]
	\nonumber
	\\
	&&
	+ f_{B_s} F_{ann}^{LR,i}\left[a_5  -\frac{1}{2}a_7 \right]
	-f_{B_s} F_{ann}^{SP,i}\left[a_{6} -\frac{1}{2}a_{8}\right]
	- M_{ann}^{LR,i}\left[C_5-\frac{1}{2}C_{7}\right]
	\nonumber
	\\
	&&
	+ M_{ann}^{LL,i}\left[C_3+C_{4}-\frac{1}{2}C_9
	-\frac{1}{2}C_{10}\right] +M_{ann}^{SP,i}
	\left[C_6-\frac{1}{2}C_{8}\right]\Bigg\},
\end{eqnarray}

\begin{eqnarray}
	\sqrt{2}A^i(\bar B_{s}^0\to\rho^{0}( \rho^{0} \rightarrow K^{+}K^{-})\phi(\phi  \rightarrow K^{+}K^{-})  )
	&=& \frac{G_Fg_{\rho}\epsilon(\lambda)\left(p_1-p_2\right)
		g_{\phi}\epsilon(\gamma)\left(q_1-q_2\right)  }{\sqrt{2}s_{\rho} s_\phi}
\nonumber
	\\
	&&
\times \Bigg \{ V_{ub}V_{us}^{*} \bigg [ f_{\rho} F_{
		B_s\to \phi}^{LL,i} \left(a_{2}\right)+ M_{ B_s\to
		\phi}^{LL,i}\left(C_{2}\right)\bigg]\nonumber
	\\
	&& -  V_{tb}V_{ts}^{*}\bigg [
	f_{\rho}F_{B_s\to \phi}^{LL,i}\frac{3}{2}\left(a_{9}+a_{7}
	\right) +M_{ B_s\to \phi}^{LL,i} \frac{3}{2}\left(C_{10}\right)
	-M_{B_s\to \phi}^{SP,i}\frac{3}{2}\left(C_{8}\right) \bigg] \Bigg\} ,
	\nonumber\\
\end{eqnarray}

\begin{eqnarray}
 \sqrt 2A^i(\bar B_{s}^0\to \omega( \omega \rightarrow K^{+}K^{-})\phi (\phi  \rightarrow K^{+}K^{-}))
&=& \frac{G_Fg_{\omega}\epsilon(\lambda)\left(p_1-p_2\right)
	g_{\phi} \epsilon(\gamma)\left(q_1-q_2\right) }{\sqrt{2}s_{\omega} s_\phi}
\nonumber
	\\
	&&\times\Bigg\{ V_{ub}V_{us}^{*} \bigg[f_\omega F_{B_s\to
	\phi}^{LL,i}\left(a_{2}\right) + M_{B_s\to
	\phi}^{LL,i}\left(C_{2}\right)\bigg] \nonumber
\\
&&-  V_{tb}V_{ts}^{*} \bigg[ f_\omega F_{B_s\to \phi}^{LL,i}\left(2a_3
+ 2a_5 +\frac{1}{2}a_7+\frac{1}{2}a_9
\right)\nonumber
\\
&&+ M_{B_s\to \phi}^{LL,i} \left(2C_4+\frac{1}{2}C_{10}\right)
-M_{B_s\to \phi}^{SP,i}\left(2C_6+\frac{1}{2}C_8\right)\bigg ]\Bigg\},
\end{eqnarray}

\begin{eqnarray}
	\sqrt{2}A^i(\bar B_{s}^0\to\rho^{0}( \rho^{0} \rightarrow K^{+}K^{-})\rho^{0}( \rho^{0} \rightarrow K^{+}K^{-})  ) &=&
	\frac{G_F g_{\rho}\epsilon(\lambda)\left(p_1-p_2\right)
		g_{\rho}\epsilon(\gamma)\left(q_1-q_2\right) }{\sqrt{2}s_{\rho} s_\rho}
\nonumber
	\\
	&&\times\Bigg\{ V_{ub}V_{us}^{*} \bigg [ f_{B_s}
	F_{ann}^{LL,i}\left(a_{2}\right)
	+ M_{ann}^{LL,i}(C_2)\bigg] \nonumber \\
	&&-   V_{tb}V_{ts}^{*}\bigg [ f_{B_s}
	F_{ann}^{LL,i}\left(2a_{3} +\frac{1}{2}a_{9} \right) +f_{B_s}
	F_{ann}^{LR,i}\left(2a_{5} +\frac{1}{2}a_{7} \right)
	\nonumber\\
	&&+M_{ann}^{LL,i}\left(2C_{4}+\frac{1}{2}C_{10}\right)+M_{ann}^{SP,i}\left(2C_{6}+\frac{1}{2}C_{8}\right)\bigg ]\Bigg\},
\end{eqnarray}
\begin{eqnarray}
	\sqrt 2 A^i(\bar B_{s}^0\to \omega( \omega \rightarrow K^{+}K^{-})\omega ( \omega \rightarrow K^{+}K^{-}) )&=&
	\frac{G_Fg_{\omega}\epsilon(\lambda)\left(p_1-p_2\right)
		g_{\omega}\epsilon(\gamma)\left(q_1-q_2\right) }{\sqrt{2}s_{\omega} s_\omega}
 \nonumber\\
	&&\times \Bigg\{V_{ub}V_{us}^{*}\bigg [ f_{B_s} F_{ann}^{LL,i}\left(a_{2}\right) +
	M_{ann}^{LL,i}(C_{2}) \bigg]\nonumber
	\\
	&&- V_{tb}V_{ts}^{*}\bigg[ f_{B_s} F_{ann}^{LL,i}\left(2a_3
	+\frac{1}{2}a_9 \right) +f_{B_s} F_{ann}^{LR,i}\left(2a_5
	+\frac{1}{2}a_7 \right)
	\nonumber\\
	&&+ M_{ann}^{LL,i}\left(2C_4+\frac{1}{2}C_{10}\right)+M_{ann}^{SP,i}\left(2C_6+\frac{1}{2}C_8\right)\bigg]\Bigg\},
\end{eqnarray}
\begin{eqnarray}
	2A^i(\bar B_{s}^0 \to \rho^{0}( \rho^{0} \rightarrow K^{+}K^{-})\omega ( \omega \rightarrow K^{+}K^{-})  ) &=& \frac{G_F g_{\rho}\epsilon(\lambda)\left(p_1-p_2\right)
		g_{\omega}\epsilon(\gamma)\left(q_1-q_2\right) }{\sqrt{2}s_{\rho} s_\omega}\Bigg\{
	V_{ub}V_{us}^{*}
	\bigg[ f_{B_s}F_{ann}^{LL,i}\left(a_{2}\right)\nonumber \\
	&&+  M_{ann}^{LL,i}(C_{2}) \bigg] -  V_{tb}V_{ts}^{*}\bigg[ f_{B_s}
	F_{ann}^{LL,i}\left(\frac{3}{2}a_9 \right)  \nonumber
	 + f_{B_s}
	F_{ann}^{LR,i}\left(\frac{3}{2}a_7 \right)	\\
	&&+M_{ann}^{LL,i}\left(\frac{3}{2}C_{10}\right)+M_{ann}^{SP,i}\left(\frac{3}{2}C_8\right)\bigg] +\left [\rho^0
	\leftrightarrow \omega\right ]\Bigg\}.
\end{eqnarray}
$C_i$ ($a_i$)  is Wilson coefficient (associated Wilson coefficient). $G_F$ stands for the Fermi constant and $f_{\phi}$ refers to the decay constants of $\phi$ \cite{Li:2006jv}.
$F_{\bar B_{s} \rightarrow \phi}^{L L}$ and $M_{\bar B_{s} \rightarrow \phi}^{L L}$ are associated with the emission graphs
encompass both factorable and non-factorable components.
The factors $F_{a n n}^{L L}$ and $M_{a n n}^{L L}$, which arise from the annihilation graphs, consist of both factorable and non-factorable contributions.
$LL$, $LR$, and $SP$ correspond to three flow structures \cite{Ali:2007ff}.

The parameters of  $V_{tb}$, $V_{t s}$, $V_{u b}$ and $ V_{u s}$ in the above equation are derived from the Cabibbo-Kobayashi-Maskawa (CKM) matrix element in the SM \cite{Cabibbo:1963yz,Kobayashi:1973}.
 $V_{t b} V_{t s}^{*}=\lambda$, $V_{u b} V_{u s}^{*}=A \lambda^{4}(\rho-i \eta)$, $V_{u b} V_{u d}^{*}=A \lambda^{3}(\rho-i \eta)(1-\frac{\lambda^{2}}{2})$, $V_{t b} V_{t d}^{*}=A \lambda^{3}(1-\rho+i \eta)$
can be obtained from the Wolfenstein parameterization.
The latest values for the parameters in the CKM matrix are $\lambda=0.22650\pm 0.00048$, $A=0.790_{-0.012}^{+0.017}$, $\bar{\rho}=0.141_{-0.017}^{+0.016}$, and $ \bar{\eta}=0.357\pm0.011$,
where  $\bar{\rho}=\rho\left(1-\frac{\lambda^{2}}{2}\right)$ and $ \bar{\eta}=\eta\left(1-\frac{\lambda^{2}}{2}\right)$  \cite{CKM}.
The physical parameters of input parameters and wave functions are arise from PDG \cite{ParticleDataGroup:2022pth, wol} .

\subsection{ Analysis of numerical results}

\begin{table}[h]
{\renewcommand
\scalebox{12}
\centering %
\renewcommand{\arraystretch}{2}
\setlength{\tabcolsep}{10mm}{
\begin{center}
\caption{	
	The peak integrated values of  $\mathrm{A}^{\Omega} _{\mathrm{CP}}$  for process $\bar B_{s} \rightarrow  K^{+} K^{-}K^{+} K^{-}$ in different resonance ranges.}
\begin{tabular}{ ccc  }
\hline
$\sqrt{s}$ $(GeV)$    & $\mathrm{0.99-1.05}$   &
\\ \hline
$\bar B_{s} \rightarrow V_1(\rho- \phi-\omega)V_2(\rho- \phi-\omega) \rightarrow K^{+}K^{-} K^{+}K^{-} $  &$\mathrm{0.0\pm0.0\%}$&

\\ \hline
$\bar B_{s} \rightarrow V_1(\rho- \phi)V_2(\rho- \phi) \rightarrow K^{+} K^{-} K^{+} K^{-} $& $\mathrm{0.0\pm0.0\%}$ &
\\ \hline
$\bar B_{s} \rightarrow V_1( \omega-\phi)V_2( \omega-\phi) \rightarrow K^{+} K^{-} K^{+} K^{-} $& $\mathrm{0.0\pm0.0\%}$ &
\\ \hline
$\bar B_{s} \rightarrow V_1(\rho- \omega)V_2(\rho- \omega) \rightarrow K^{+} K^{-} K^{+} K^{-} $& $\mathrm{1.57\pm0.01\pm0.90\%}$ &
\\ \hline

$\bar B_{s} \rightarrow V_1V_2(no  \:\: \phi\phi) \rightarrow K^{+}K^{-} K^{+}K^{-}  $       & $\mathrm{-21.47\pm0.61\pm5.70\%}$   &
\\
\hline

$\bar B_{s} \rightarrow V_1(\phi)V_2(\rho-\omega-\phi) \rightarrow K^{+}K^{-} K^{+}K^{-}  $       & $\mathrm{0.0\pm0.0\%}$   &
\\ \hline
$\bar B_{s} \rightarrow V_1(\phi)V_2(\rho-\omega) \rightarrow K^{+}K^{-} K^{+}K^{-}  $       & $\mathrm{-76.59\pm3.26\pm9.01\%}$   &
\\ \hline
$\bar B_{s} \rightarrow V_1(\phi)V_2(\rho-\phi) \rightarrow K^{+}K^{-} K^{+}K^{-}  $       & $\mathrm{0.0\pm0.0\%}$   &
\\ \hline
$\bar B_{s} \rightarrow V_1(\phi)V_2(\omega-\phi) \rightarrow K^{+}K^{-} K^{+}K^{-}  $       & $\mathrm{0.0\pm0.0\%}$   &
\\ \hline
\end{tabular}
\end{center}}}
\end{table}

 To enhance our comprehension of local CP violation and provide a theoretical framework for future experiments, we have conducted an analysis on the localized integration of CP violation for the  process of $\bar{B}_{s} \rightarrow K^{+} K^{-} K^{+} K^{-}$. The corresponding numerical results are presented in Table I. The first uncertainty here corresponds to the CKM parameters, the second arises from PQCD method.
The integration interval (0.99 GeV-1.05 GeV) from Table I corresponds to the resonance interference associated with the decay process $V \rightarrow K^{+}K^{-}$, which is defined within a narrow range of ${\pm0.06}$ GeV centered around the invariant mass of $K^{+}K^{-}$, denoted as $\sqrt{s}$=1.02 GeV.

The second row in Table I represents the local integrals of the $\bar{B}_{s}^{0} \rightarrow K^{+} K^{-} K^{+} K^{-}$ decay process involving three particles ($\phi-\rho-\omega$) for $V_1$ or $V_2$ vector meson
within a specified integration range. From the third to fifth rows, CP violation resulting from the decay $\bar{B}_{s}^{0}\rightarrow  K^{+} K^{-} K^{+} K^{-}$ is presented under different conditions of mixing between any two vector particles. For the case of $\rho$-$\omega$ mixing, the calculated local integral for this decay process within an energy range of 0.99 GeV-1.05 GeV yields a central value of $1.57\%$. However, for both $\omega$-$\phi$ and $\rho$-$\phi$ mixings, no significant CP violation is observed. If neither $V_1$ nor $V_2$ simultaneously refers to the $\phi$ meson, the decay process $\bar B_{s} \rightarrow V_1V_2(no  \:\: \phi\phi) \rightarrow K^{+}K^{-} K^{+}K^{-}$ exhibits a relatively significant magnitude of CP violation with a central value of $-21.47\%$, as shown in Table I, which process cannot be controlled experimentally.

From the seventh row to the tenth row, we assign $V_1$ to represent the $\phi$ meson and $V_2$ to denote the mixing of different vector mesons, which quantifies CP violation. It is observed that most decay processes exhibit no evidence of CP violation. However, if $V_2$ corresponds to the $\rho-\omega$ mixing, significant CP violation can be obtained with a large central value of $-76.59 \%$. This phenomenon is associated with the process $\bar B_{s} \rightarrow V_1(\phi)V_2(\rho-\omega) \rightarrow K^{+}K^{-} K^{+}K^{-}$ as presented in Table I.

 The penguin-dominant decay process $\bar{B}_{s}^{0}\rightarrow \phi\phi \rightarrow K^{+} K^{-} K^{+} K^{-}$ does not exhibit a phase that triggers CP violation through an intermediate state of $\phi\phi$ in the PQCD approach. However, due to the new phase introduced from the mixing of $\rho$ and $\omega$, CP violations are observed in the decay processes of $\bar{B}_{s} \rightarrow V_1(\rho- \omega)V_2(\rho- \omega) \rightarrow K^{+} K^{-} K^{+} K^{-}$ and $\bar{B}_{s} \rightarrow V_1(\phi)V_2(\rho-\omega) \rightarrow K^{+}K^{-} K^{+}K^{-}$. The detection of the predicted CP violation in the decay process of the $\bar{B}_{s}^{0}$ meson can be achieved by reconstructing $\phi$, $\omega$, and $\rho$ mesons from the invariant mass of $K^{+} K^{-}$ meson pairs within the resonance region during experiments.

\begin{table}[h]
{\renewcommand
\scalebox{12}
\centering %
\renewcommand{\arraystretch}{2}
\setlength{\tabcolsep}{2.5mm}{
\begin{center}
\caption{The $\mathrm{A} _{\mathrm{CP}}$ value of the $\bar B_{s} \rightarrow K^{-} K^{+}K^{-} K^{+}$ decay process with different intermediate states.}
\begin{tabular}{ cc }
 \hline
 $\mathrm{A} _{\mathrm{CP}}$ ($\bar B_{s} \rightarrow \phi \phi \rightarrow K^{+} K^{-} K^{+} K^{-} $ )  =$\mathrm{0\pm0\pm0\%}$    & \;\;\;\;$\mathrm{A} _{\mathrm{CP}}$ ($\bar B_{s} \rightarrow \rho \omega \rightarrow K^{+} K^{-} K^{+} K^{-} $ )  =$\mathrm{-1.6\pm0.01\pm2.3\%}$
 \\
\hline
 $\mathrm{A} _{\mathrm{CP}}$ ($\bar B_{s} \rightarrow \rho \rho \rightarrow K^{+} K^{-} K^{+} K^{-} $)=$\mathrm{1.2\pm0.1\pm4.1\%}$  & $\mathrm{A} _{\mathrm{CP}}$ ($\bar B_{s} \rightarrow \rho \phi \rightarrow K^{+} K^{-} K^{+} K^{-} $ )  =$\mathrm{7.7\pm0.1\pm2.0\%}$
\\ \hline
 $\mathrm{A} _{\mathrm{CP}}$ ($\bar B_{s} \rightarrow \omega \omega \rightarrow K^{+} K^{-} K^{+} K^{-} $)=$\mathrm{3.8\pm0.0\pm1.2\%}$ & \;\;\;\;\;$\mathrm{A} _{\mathrm{CP}}$ ($\bar B_{s} \rightarrow  \omega \phi \rightarrow K^{+} K^{-} K^{+} K^{-} $ )  =$\mathrm{-4.2\pm0.2\pm8.6\%}$
\\ \hline
\end{tabular}
\end{center}}}
\end{table}

In Table II, we present the results of the four-body decay process independent of any mixing effects from intermediate
states, which  provides valuable reference for the experiments.
The calculation results indicate that the $\bar B_{s} \rightarrow \phi \phi \rightarrow K^{+} K^{-} K^{+} K^{-} $ decay process has an $\mathrm{A} _{\mathrm{CP}}$ value of $\mathrm{0\%}$, which is to be expected.
The central values of $\mathrm{A_{CP}}$ for the decay processes $\bar B_{s} \rightarrow \rho \rho \rightarrow K^{+} K^{-} K^{+} K^{-}$ and $\bar B_{s} \rightarrow \omega \omega \rightarrow K^{+} K^{-} K^{+} K^{-}$ are $1.2\%$ and $3.8\%$, respectively. 
In the decay process $\bar B_{s} \rightarrow \rho \omega \rightarrow K^{+} K^{-} K^{+} K^{-}$, the measured central value of $\mathrm{A}_{\mathrm{CP}}$ is $-1.6\%$. Similarly, for the decay process $\bar B_{s} \rightarrow \rho \phi \rightarrow K^{+} K^{-} K^{+} K^{-}$, the central value of $\mathrm{A}_{\mathrm{CP}}$ is $7.7\%$. Notably, a CP violation with a magnitude of $-4.2\%$ is observed in the decay process $\bar B_{s} \rightarrow \omega \phi \rightarrow K^{+} K^{-} K^{+} K^{-}$.

The branching ratios and CP violation are presented for the two-body decay process of $B_{s} \rightarrow VV$ within the framework of PQCD \cite{Ali:2007ff}. In PQCD, the Gegenbauer moments in the two-meson distribution amplitudes of the $KK$ system are improved by fitting the PQCD factorization formulas to the measured branching ratios of the four-body decay $B_{s} \rightarrow \phi\phi \rightarrow K^{+}K^{-}K^{+}K^{-}$. The two-meson distribution amplitude describes the collinear motion of the two mesons. The hard kernel, which captures the strong and electroweak interactions, can be derived from the corresponding two-body decays. Since no tree-level operators contribute to the four-body decay $B_{s} \rightarrow \phi\phi \rightarrow K^{+}K^{-}K^{+}K^{-}$, there is no direct CP violation \cite{PRD105-093001-2022}. 
However, the amplitudes of the four-body decay can be calculated within the PQCD framework by utilizing the Breit-Wigner formula for the intermediate state and the amplitude derived from the decay of the intermediate particle into the final states. Within our PQCD framework for the four-body decay process, the CP violations from our calculations are consistent with those from the two-body decay process, which is attributed to the strong decay of hadrons, as shown in Table II.

It has been determined that the impact of mixing parameter errors on local CP violation is negligible, accounting for less than one-thousandth of the overall CP violation. Therefore, we have omitted the specific numerical value associated with this influence in this context.
It is well known that CP violation arises from the weak phase provided by CKM as well as from the strong phase determined by  the ratio of the penguin diagram and tree diagram contributions.
The values of CKM elements are obtained through experiments, while there are some errors. In this work, we introduce error analysis because there is uncertainty that needs to be considered when calculating amplitudes using perturbative QCD methods.
The primary source of error comes from the uncertainty range associated with CKM parameters. The discrepancy in CP violation results calculated using the maximum and minimum values of CKM elements is relatively small when compared to the results obtained with intermediate values.
 The second source of error arises from the strong interaction parameters, including meson decay constants, form factors, and uncertainties in the wave functions of relevant mesons.

 We investigate the impact of intermediate vector particle interference on CP violation in the decay process $B_s\rightarrow VV\rightarrow K^+K^{-} K^+K^{-}$. Our findings suggest that there is a potential for significant CP violation in the decay channels of $\bar{B}_s$ mesons, as indicated by the $\phi-\rho-\omega$ mixing prediction.
The numbers of required $B_s \bar{B}_s$ pairs for observing $CP$ violation
depend on the magnitude of CP violation and the branching ratio of heavy hadron decays. For one (three) standard deviation signature,
the number of $B_s \bar{B}_s$ pairs is
  \cite{Du:1986ai,Lyons,Eadie,Guo:2008zzh,Lu:2013xea}:
\begin{eqnarray}	
		N_{B_{s} \bar{B}_{s}} \sim \frac{1}{B R A_{C P}^{2}}\left(1-A_{C P}^{2}\right) \sim \left(\frac{9}{B R A_{C P}^2}\left(1-A_{C P}^{2}\right)\right).
\end{eqnarray}
 where  BR represents the branching ratio of $\bar{B}_s\rightarrow VV$  and $A_{CP}$ is the CP violation value.

\begin{table}[h]
	{\renewcommand
		\scalebox{12}
		\centering %
		\renewcommand{\arraystretch}{2}
		\setlength{\tabcolsep}{1.2mm}{
			\begin{center}
				\caption{ The numbers of  $B_s \bar{B}_s$ pairs  required to observe CP violation in the $\bar{B}_s$ decay channel at  one (three) standard deviation signature.}
				\begin{tabular}{ cccc }
					\hline
					Decay& Numbers of $B_s \bar{B}_s$  & \;\;\;\;Decay& Numbers of $B_s \bar{B}_s$
					\\
					\hline
					$\bar B_{s} \rightarrow \phi \phi \rightarrow K^{+} K^{-} K^{+} K^{-} $  & $-$ \;\;&\;  $\bar B_{s} \rightarrow \rho \phi \rightarrow K^{+} K^{-} K^{+} K^{-} $
					$$&$7.29 (65.61 )\times10^8$
					\\
					$\bar B_{s} \rightarrow \rho \rho \rightarrow K^{+} K^{-} K^{+} K^{-} $ &$1.46(13.17)\times 10^{8}$   $$&$\bar B_{s} \rightarrow \rho \omega \rightarrow K^{+} K^{-} K^{+} K^{-} $   & $5.58 (50.21)\times 10^{11}$
					\\
					$\bar B_{s} \rightarrow \omega \omega \rightarrow K^{+} K^{-} K^{+} K^{-} $ &$1.77(15.96)\times 10^{9}$ &  $\bar B_{s} \rightarrow  \omega \phi \rightarrow K^{+} K^{-} K^{+} K^{-} $   &$2.96(26.72)\times 10^{7}$
					\\
					$\bar B_{s} \rightarrow V_1(\rho$ - $\omega)
					V_2(\rho$ - $\omega) \rightarrow K^{+} K^{-} K^{+} K^{-} $ &$1.33(11.98)\times 10^{10}$ & 					
					
					\\
						$\bar B_{s} \rightarrow V_1(\phi)V_2(\rho$ - $\omega) \rightarrow K^{+} K^{-} K^{+} K^{-} $ &$1.81(16.26)\times 10^{6}$
					\\
					\hline
				\end{tabular}
	\end{center}}}
\end{table}

The numbers of $B_s \bar{B}_s$ pairs for observing $CP$ violation are presented in Table III.
One can find that the number of required $B_s \bar{B}_s$ pairs is $10^6$ $\sim $ $10^{11}$ in order to observe significant CP violation.
The Large Hadron Collider (LHC) is a proton-proton collider built
at CERN with the center-of-mass energy 14 TeV and luminosity
$L=10^{34}cm^{-2}s^{-1}$. The $b\bar{b}$ production cross section
 is huge and of the order of $500\mu b$, providing $0.5 \times 10^{12}$
bottom events per year \cite{Schopper2005}.
If a nominal annual integrated luminosity of
$L_{int}=2 fb^{-1}$ and a $b\bar{b}$ production cross
section of $\sigma_{b\bar{b}}=500\mu b$.
The probability for a $\bar{b}$-quark to hadronize
into a hadron is assumed to be $f_{B_{s}}=10\%$ for $B_{s}$.
The factor 2 takes into account the production of both
$b$- and $\bar{b}-$ hadrons \cite{CERN2003-030}.
Ignoring small asymmetry between the numbers of b-hadrons and
those of their antiparticles in
the Lund string fragmentation model and
the intrinsic quark model,
the LHC can provide
about $10^{10}$ of $B_s \bar{B}_s$ pairs \cite{CERN2000-004, Norrbin99}.

The LHCb experiment has collected data of B mesons about  $1fb^{-1}$ at $\sqrt{s}=1$ TeV, $2fb^{-1}$ at 8 TeV, and close to  $5.9 fb^{-1}$ at 13 TeV during Runs 1 and 2.
The ATLAS and CMS at the LHC have collected each about  $5 fb^{-1}$ at
$\sqrt{s}=7$ TeV, $20 fb^{-1}$ at 8 TeV, and about $150 fb^{-1}$ at 13 TeV
during Runs 1 and 2.
With these data, we are entering to regime of precision physics
even for many rare decay \cite{ParticleDataGroup:2022pth}.
 Flavour physics can potentially be studied in the High-Luminosity phase of the Large Hadron Collider (HL-LHC) and its possible upgrade to a 27 TeV proton collider, known as the High-Energy LHC (HE-LHC), in the future.  The CP violation associated  with flavour physics could be
  measured with higher precision.
 The prospective experimental sensitivities for the HL-LHC assume
 $3000 fb^{-1}$ recorded by ATLAS and CMS, and $300 fb^{-1}$ recorded by a proposed Upgrade II of LHCb \cite{HLC867}.

Therefore, it is possible to observe the predicted CP violation by collecting a range of $10^6$ to $10^{11}$ pairs of $B_s \bar{B}_s$ at the LHC experiment or future HL-LHC and HE-LHC experiments.

\section{Summary and conclusion}

Recently, in experimental studies, the CDF \cite{prl107-261802} and LHCb Collaborations \cite{plb713-369,LHCb:2013xyz,prd90-052011} have reported precise measurements of CP violation for the $\bar{B}_{s}^{0}\rightarrow \phi\phi \rightarrow K^{+} K^{-} K^{+} K^{-}$ decay mode, revealing no significant deviations from the predictions of the SM. Notably, no evidence of CP violation has been observed in the $\bar{B}_{s}^{0}\rightarrow \phi\phi \rightarrow K^{+} K^{-} K^{+} K^{-}$ process.

We investigate the CP violation in the four-body decay process of $\bar{B}_{s}^{0} \rightarrow K^{+} K^{-} K^{+} K^{-}$, involving intermediate vector mesons and their mixing. Notably, significant CP violation is observed for specific two-vector meson intermediate states. Additionally, a substantial amount of CP violation arises from vector mixing when the invariant mass of $K^{+}K^{-}$ is localized within a certain range.
The required numbers of $B_s \bar{B}_s$ pairs for observing predicted $CP$ violation in experiments at the LHC are also presented. The detection of predicted CP violation in the decay process of $\bar{B}_{s}^{0}$ meson can be achieved by reconstructing $\phi$, $\omega$ and $\rho$ mesons from the invariant mass of $K^{+} K^{-}$ meson pairs within the resonance region during experiments. This study has potential implications for future detection by the LHC experiment.

 \section*{Acknowledgements}
 We express our sincere gratitude to Professor Yue-Hong Xie for his  insightful discussions regarding the LHC experiment.
 This work was supported by  Natural Science Foundation of Henan (Project No. 232300420115).


\end{spacing}
\end{document}